\documentclass{moriond}

\def\Journal#1#2#3#4{{#1} {\bf #2}, #3 (#4)}

\def\PRL{\em Phys. Rev. Lett.}

\def\be{\begin{equation}}
\def\ee{\end{equation}}
\def\bea{\begin{eqnarray}}
\def\eea{\end{eqnarray}}

\usepackage{xspace}
\usepackage{amsmath}
\usepackage{lineno}

\def \pp{\ensuremath{pp}\xspace}
\def \pHe{\ensuremath{p\text{He}}\xspace}
\def \pPb{\ensuremath{p\text{Pb}}\xspace}

\def \PbPb{\ensuremath{\text{PbPb}}\xspace}

\def \ie{\textit{i.e.}\xspace}
\def \pt{\ensuremath{p_\text{T}}\xspace}
\def \snn{\ensuremath{\sqrt{s_\text{NN}}}\xspace}
\newcommand{\decay}[2]{\ensuremath{#1\!\to #2}\xspace}

\def\pion   {{\ensuremath{\mathrm{\pi}}}\xspace}
\def\piz    {{\ensuremath{\pion^0}}\xspace}
\def\pip    {{\ensuremath{\pion^+}}\xspace}
\def\pim    {{\ensuremath{\pion^-}}\xspace}
\def\kaon   {{\ensuremath{\mathrm{K}}}\xspace}

\def\Km    {{\ensuremath{\kaon^-}}\xspace}
\def\electron   {{\ensuremath{\mathrm{e}}}\xspace}
\def\en         {{\ensuremath{\electron^-}}\xspace}   
\def\ep         {{\ensuremath{\electron^+}}\xspace}
\def\muon       {{\ensuremath{\mathrm{\mu}}}\xspace}
\def\mup        {{\ensuremath{\muon^+}}\xspace}
\def\mun        {{\ensuremath{\muon^-}}\xspace} 
\def\jpsi     {{\ensuremath{{\mathrm{J}\mskip -3mu/\mskip -2mu\mathrm{\psi}}}}\xspace}
\def\Dz      {{\ensuremath{\mathrm{D}^0}}\xspace}
\def\B       {{\ensuremath{\mathrm{B}}}\xspace}
\def\Bs      {{\ensuremath{\B^0_s}}\xspace}
\def\Bz      {{\ensuremath{\B^0}}\xspace}
\def\BsBz    {{\ensuremath{\B^0_{(s)}}}\xspace}
\def\Lc      {{\ensuremath{\Lambda_c}}\xspace}

\newcommand{\offsetoverline}[2][0.18em]{\kern #1\overline{\kern -#1 #2}}%
\def\Lbar    {{\ensuremath{\offsetoverline{\Lambda}}}\xspace}
\def\Sbar    {{\ensuremath{\offsetoverline{\Sigma}^-}}\xspace}
\def\Xibarp  {{\ensuremath{\offsetoverline{\Xi}^+}}\xspace}
\def\Xibarz  {{\ensuremath{\offsetoverline{\Xi}^0}}\xspace}
\def\Omegabar    {{\ensuremath{\offsetoverline{\Omega}^+}}\xspace}

\def\Hbar    {{\ensuremath{\overline{H}}}\xspace}
\def\pbar    {{\ensuremath{\overline{p}}}\xspace}

\newcommand{\Photo}{}

\begin{document}
\vspace*{4cm}
\title{HEAVY-ION AND FIXED-TARGET PHYSICS AT LHCB}

\author{SAVERIO MARIANI, on behalf of the LHCb collaboration}

\address{Istituto Nazionale di Fisica Nucleare (INFN), Sezione di Firenze,\\ via Giovanni Sansone 1, Sesto Fiorentino (FI), Italy}

\maketitle
\abstracts{In parallel to the study of proton-proton collisions, LHCb is developing a unique heavy-ion programme and is pioneering beam-gas fixed-target physics at the CERN LHC. In this document, a selection of some recent results from both programmes is presented.}

\section{Introduction}
\begin{figure}[tb]
	\centering
	\includegraphics[width=0.51\textwidth]{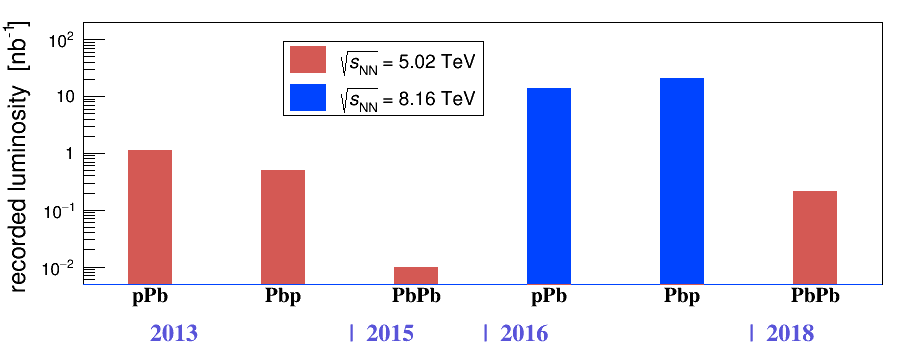}
	\includegraphics[width=0.47\textwidth]{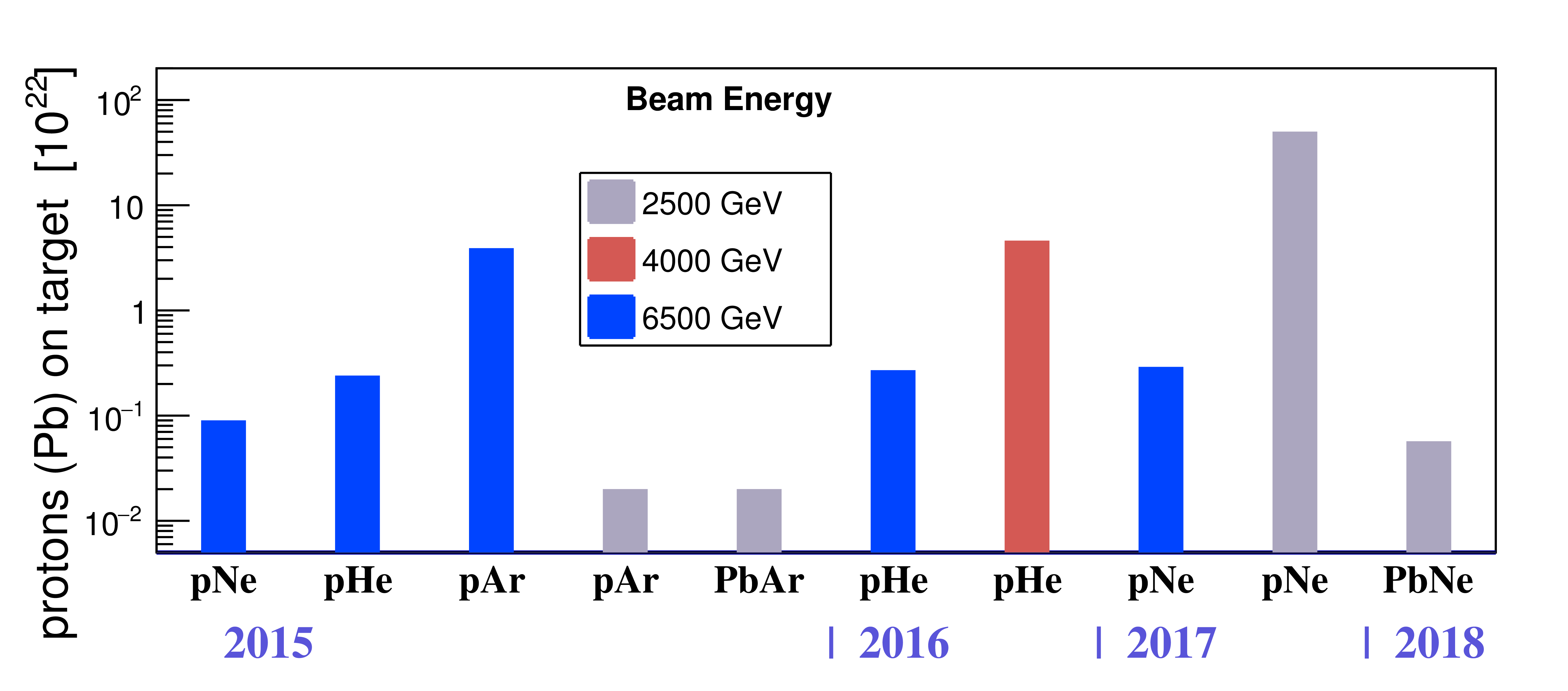}	
	\caption{Heavy-ion samples collected by LHCb in its (left) collider and (right) fixed-target mode.}
	\label{fig:samples}
\end{figure}

The LHCb experiment\,\cite{LHCb-DP-2008-001} at CERN, originally designed for heavy flavour physics studies in proton-proton (\pp) collisions, is a single-arm spectrometer instrumenting the pseudorapidity region $\eta \in [2,5]$. Leveraging on the forward acceptance, complementary to the other experiments operating at the CERN LHC, and on the excellent particle reconstruction and identification performance, the LHCb physics reach has progressively evolved and also embraces now a heavy-ion and a fixed-target programme. Figure~\ref{fig:samples} summarises the ion samples that have been collected in the LHC Run2, from 2015 to 2018. In collider mode, proton-lead (\pPb) collisions in the forward (backward) configuration, \ie with the proton (the lead) beam entering from the vertex detector, give access to \textit{x} values down to $\mathcal{O}(10^{-6})$ (up to $\mathcal{O}(10^{-1})$). Lead-lead (\PbPb) data were also acquired in the 60 - 100\% centrality range, limited because of the hardware saturation due to the high track density in the forward region. Such a limitation will be reduced by half already with the ongoing upgrade of the experiment and is expected to be completely removed in the future thanks to the installation of more granular tracking detectors.\\\\
By injecting gases in the LHC beam-pipe, LHCb has also operated since 2015 in fixed-target mode~\cite{CERN-THESIS-2021-313}, collecting samples with proton and lead beams impinging on gaseous targets. The high-\textit{x} and moderate $Q^2$ regime, poorly explored by previous experiments, can now be precisely probed in different collision systems. From 2022, the physics opportunities accessible to the fixed-target programme will be further extended\,\cite{LHCb-PUB-2018-015} by confining the gas in a 20-cm-long cell upstream of the LHCb nominal interaction point\,\cite{LHCb-TDR-020}. The gas pressure will be increased by up to two orders of magnitude for the same gas flow as in Run2 and heavier noble gases like krypton and xenon or non-noble species such as hydrogen, deuterium or oxygen could be injected. Owing to the separation between the beam-beam and beam-gas interaction regions, LHCb could be the only experiment operating at the same time in collider and fixed-target modes at two energy scales.

\section{The LHCb heavy-ion programme}
\begin{figure}
	\centering
	\includegraphics[width = \textwidth]{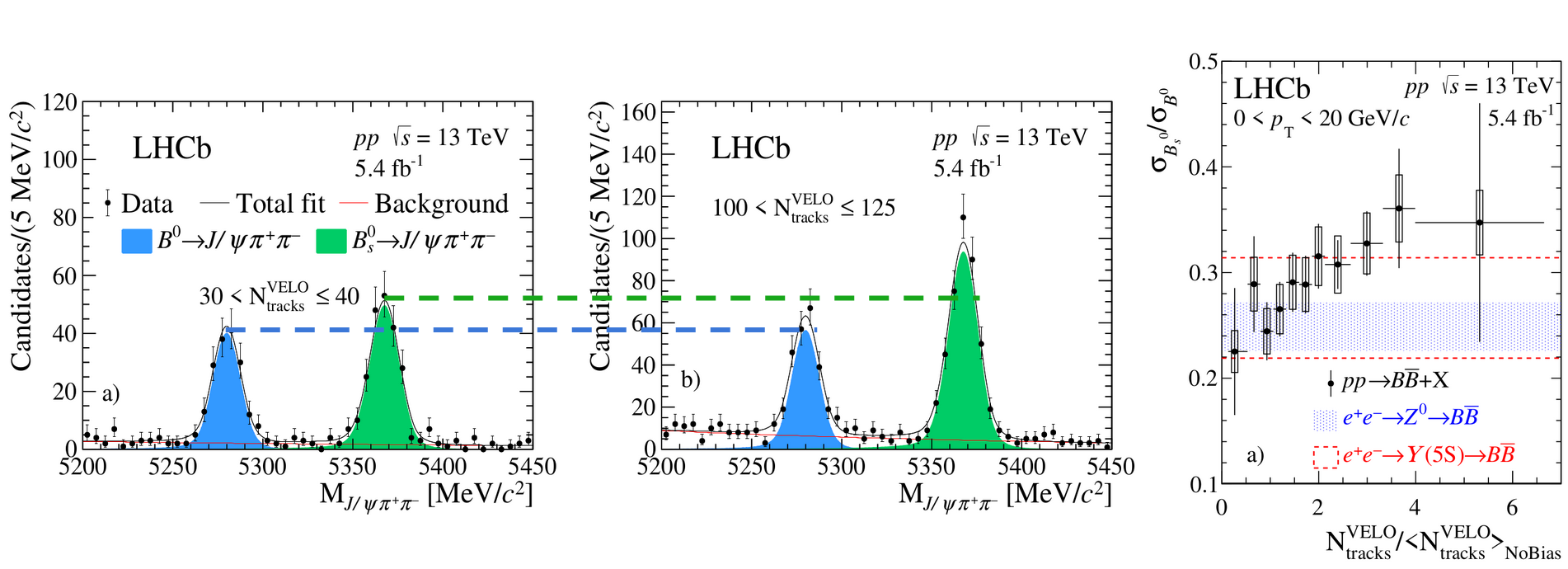}
	\caption{Measurement of the \Bs-to-\Bz production ratio in \pp collisions. The left and middle plots show the invariant mass distributions for the \jpsi\pip\pim candidates in two intervals of the number of reconstructed tracks in the VELO detector and an evident enhancement of the strangeness production can be seen. The right plot presents the result of the measurement for the cross-section ratio as a function of the multiplicity.}
	\label{fig:Bhadron}
\end{figure}
\begin{figure}
	\centering
	\includegraphics[width = \textwidth]{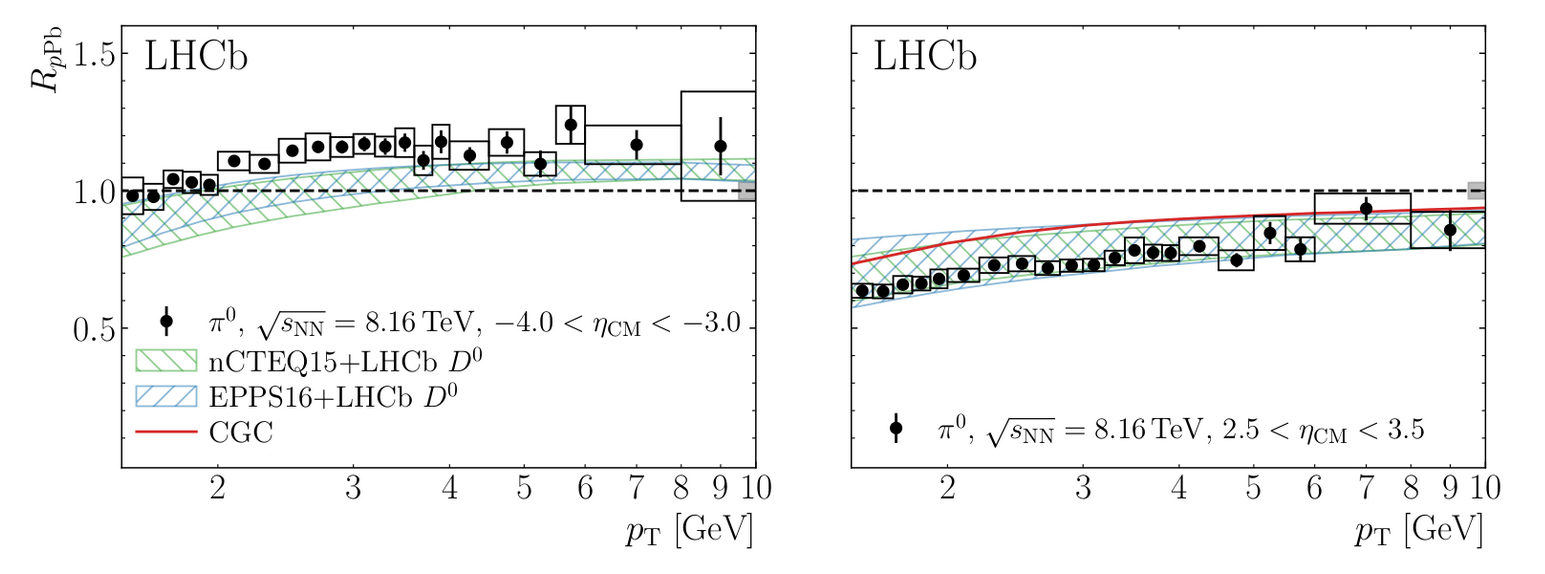}
	\caption{Measurement of the neutral pions nuclear modification factor as a function of the transverse momentum with (left) backward and (right) forward \pPb collisions compared to theoretical models.}
	\label{fig:RpPb}
\end{figure}

Leveraging on its unique characteristics, LHCb is carrying out a rich programme of heavy-ion interest. Among the most recent results, one example for collision system is discussed here.\\
The possible \textit{b} quark hadronization via coalescence is probed in \pp collisions at $\sqrt{s} = 13$~TeV by measuring the \Bs-to-\Bz production ratio\,\cite{LHCb-PAPER-2022-001} as a function of the detector multiplicity. Coalescence is expected to increase with the density of the produced particles, maximum at low transverse momentum, and can be thus efficiently addressed by LHCb. Both hadrons are reconstructed in their common decay mode \decay{\BsBz}{ (\decay{\jpsi}{\mup\mun})\pip\pim}. Left and middle  Figure~\ref{fig:Bhadron} show the final-state invariant mass distributions in two intervals of the number of tracks reconstructed in the vertex detector (VELO) and, by eye, an increase in the strangeness production is observed. The production ratio, obtained by correcting the fitted \Bs and \Bz yields by the respective efficiencies, is illustrated in the right. The measurement is found to be consistent with previous estimations from \ep\en colliders at low multiplicity and increases at larger values. Being the production ratio only dependent on the \textit{b}-quark hadronization, such a result qualitatively hints at the occurrence of hadronization mechanisms other than fragmentation.\\\\
\begin{figure}
	\centering
	\includegraphics[width = 0.48\textwidth]{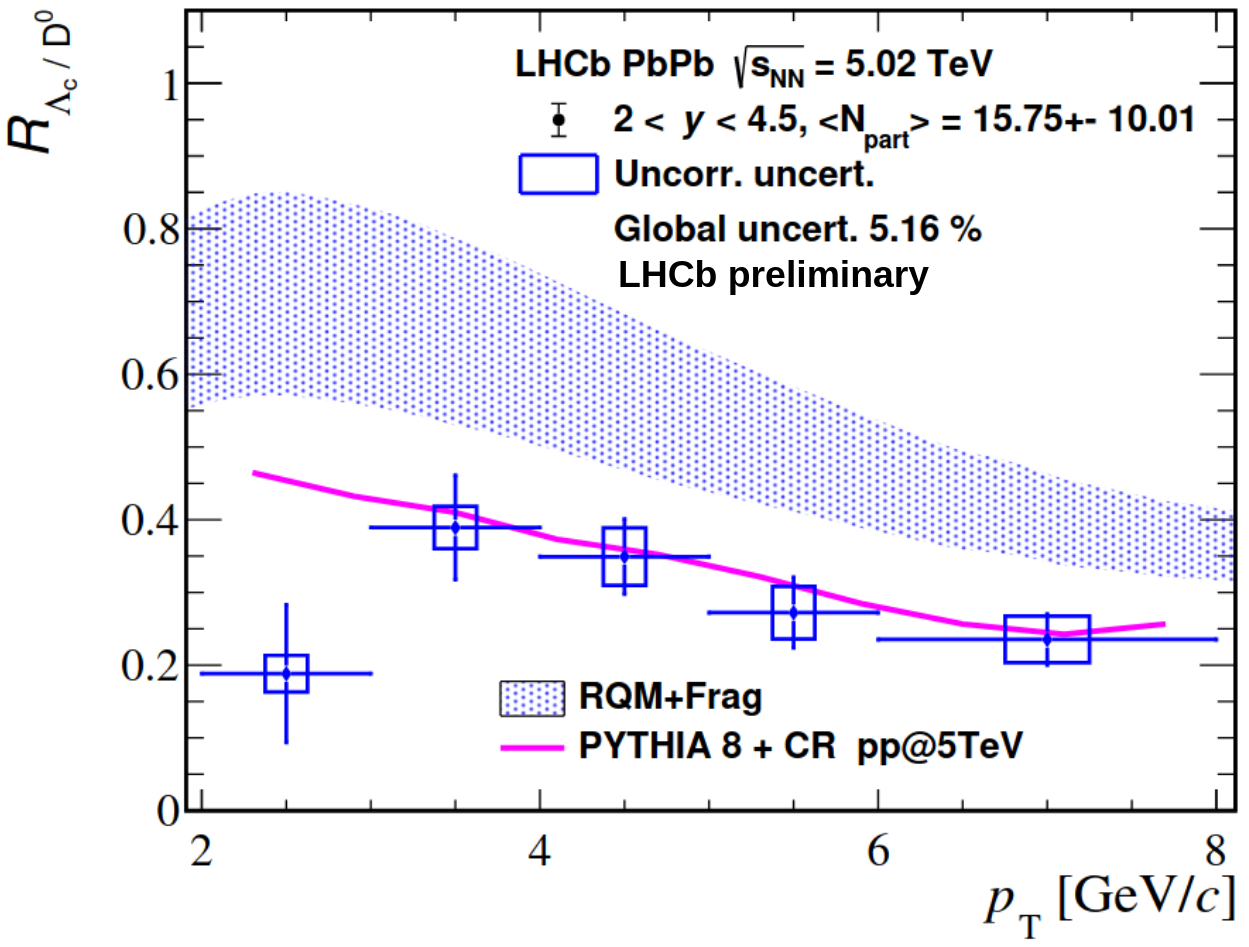}
	\includegraphics[width = 0.48\textwidth]{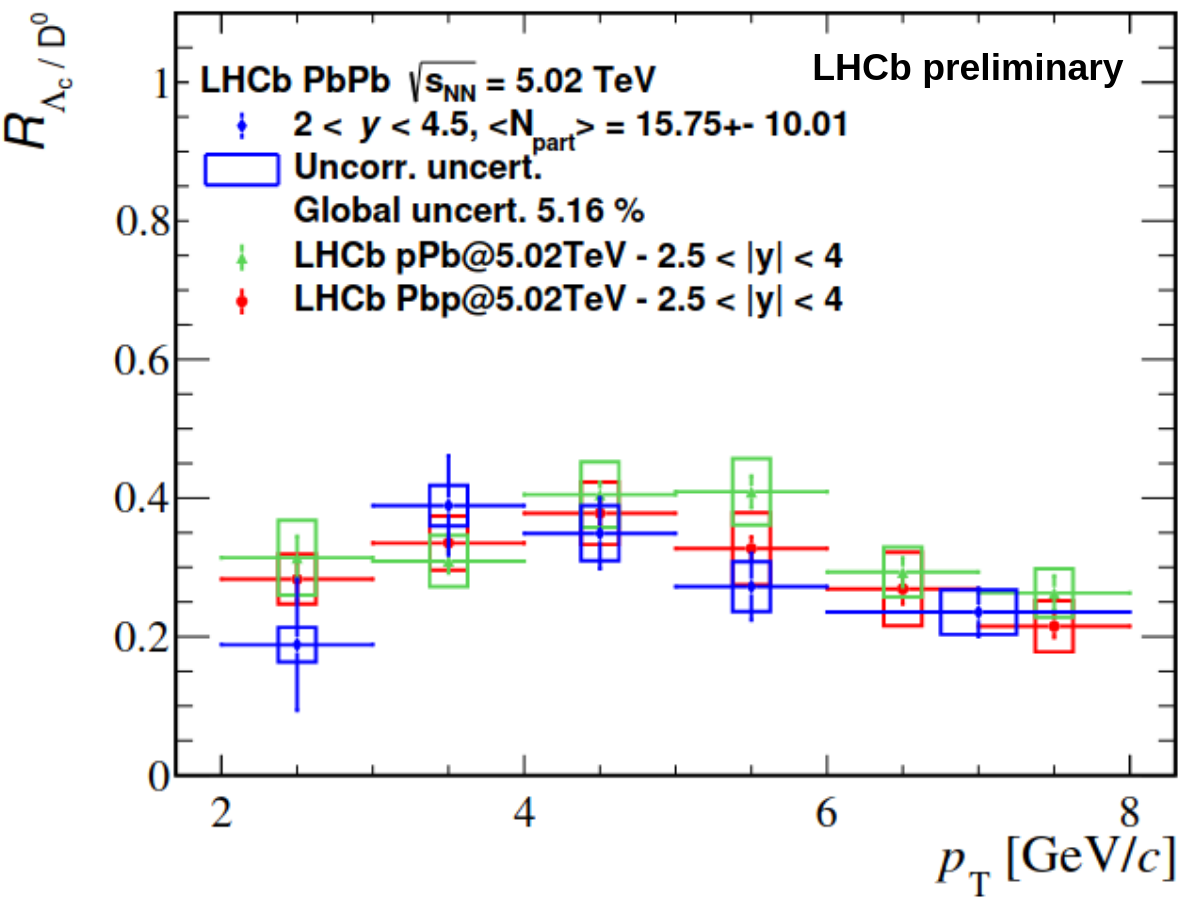}
	\caption{Measurement of the \Lc-to-\Dz ratio as a function of the transverse momentum compared to (left) theoretical models and to (right) a previous LHCb measurement with \pPb data.}
	\label{fig:LcD0_results}
\end{figure}
Hadronization of the \textit{c}-quark has been recently probed by measuring the \Lc-to-\Dz ratio in peripheral \PbPb collisions\,\cite{LHCb-PAPER-2021-046} at \snn = 5.02~TeV. The results, obtained considering the \decay{\Lc}{p\Km\pip} and \decay{\Dz}{\Km\pip} final states, are shown as a function of the transverse momentum in Figure~\ref{fig:LcD0_results} compared (left) to theoretical models and (right) to a previous \pPb LHCb measurement~\cite{LHCb-PAPER-2018-021}. While the model including the color reconnection mechanism is found to well describe the data for $\pt > 3$~GeV/c, a disagreement with the statistical hadronization one is found. The measurement with \PbPb data, flat as a function of the number of participants, confirms the previous LHCb findings and the difference with respect to the results from mid-rapidity experiments. This supports the evidence of a dependence of the \Lc-to-\Dz ratio on the rapidity.\\\\
LHCb is also providing unique inputs to the study of cold nuclear matter effects, for example with the recent measurement of the nuclear modification factor of neutral pions in forward and backward \pPb collisions\,\cite{LHCb-PAPER-2021-053}. The \decay{\piz}{\gamma\gamma} yields are measured in \pPb collisions in the (left) backward and (right) forward configurations and in reference \pp samples. The resulting nuclear modification factor, $R_{\pPb} = \dfrac{1}{208} \cdot \dfrac{\text{d}\sigma_{\pPb}/\text{d}\pt}{\text{d}\sigma_{\pp}/\text{d}\pt}$, is shown as a function of the transverse momentum in Figure~\ref{fig:RpPb} compared to nuclear PDF (nPDF) theoretical models. The results, affected by an uncertainty below 6\%, will significantly contribute to constrain the models. Forward results are found to be well described by nPDF models but overestimated from the color glass condensate one; the backward measurement indicates a small excess over nPDF expectations.

\section{The LHCb fixed-target programme}
\begin{figure}
\centering
\includegraphics[width = 0.48\textwidth]{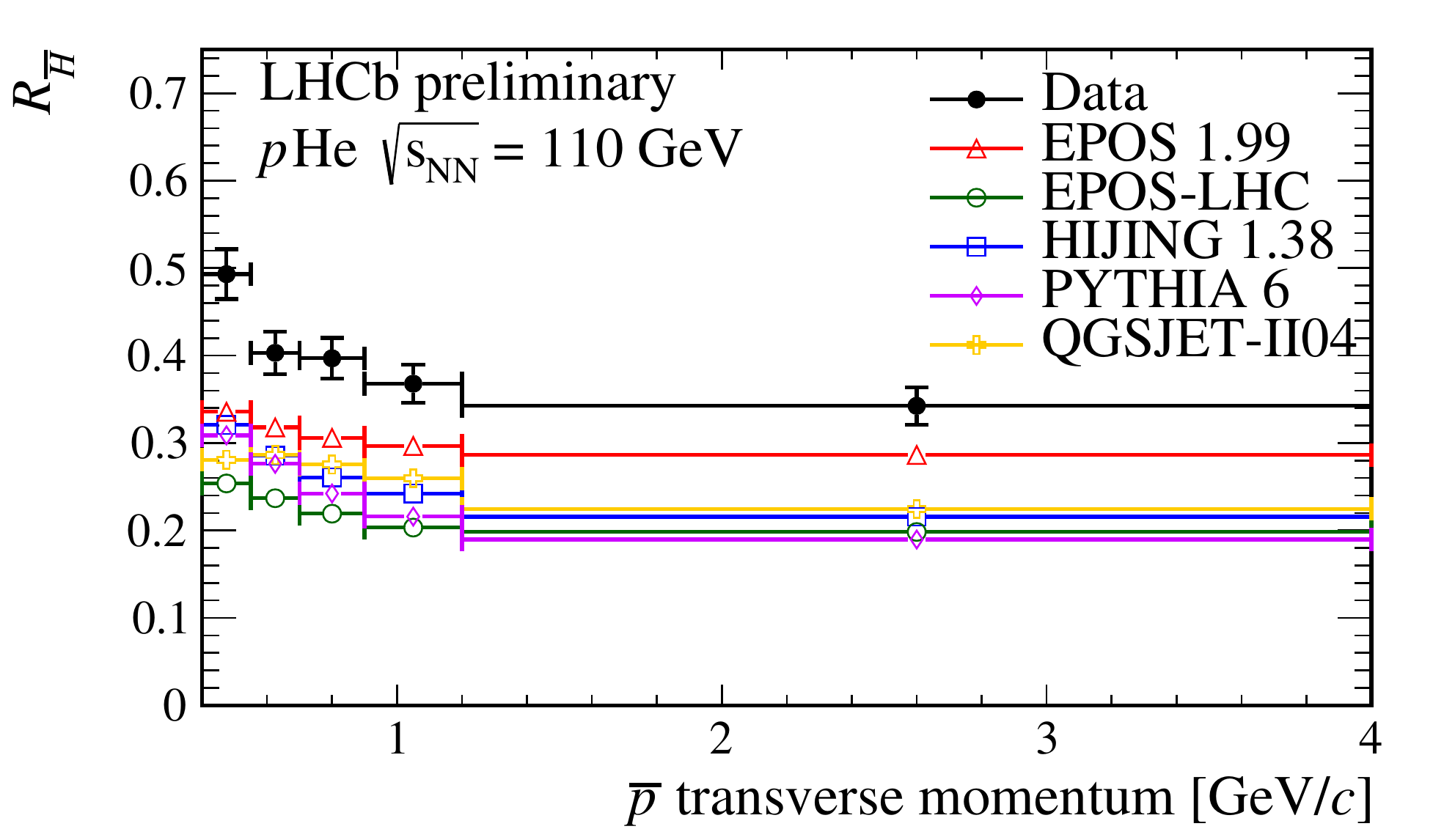}
\includegraphics[width = 0.48\textwidth]{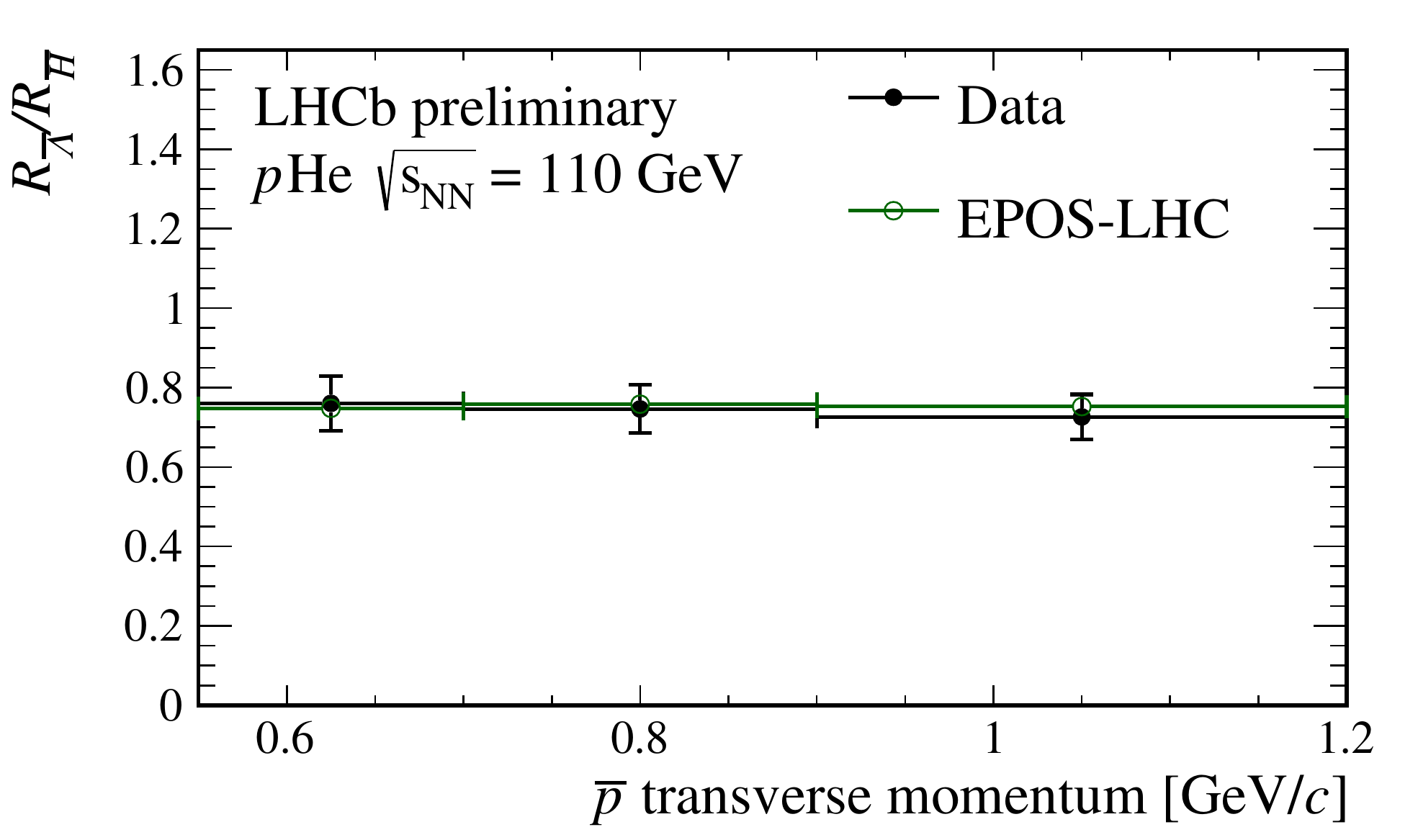}
\caption{Measurement of (left) the $R_{\Hbar}$ ratio as a function of the transverse momentum compared to the predictions of the most widely used theoretical models, all found to largely underestimate the quantity. The right plot compares the $R_{\Lbar}/R_{\Hbar}$ measurement with the EPOS-LHC prediction, expected to be more reliable since only related to the \textit{s}-quark hadronization, and an overall consistency is observed.}
\label{fig:pHe_results}
\end{figure}
A measurement motivated by cosmic rays physics has also been recently performed by fixed-target LHCb. By injecting helium, the antiproton production in \pHe collisions is measured, providing a crucial ingredient to constrain the flux of antiprotons in space originating in cosmic rays spallation on the interstellar medium, mainly composed of hydrogen and helium. The description of this process, background to dark matter decay or annihilation searches, currently limits the interpretation of the antiproton flux data collected by orbital experiments. After the first measurement ever of prompt antiproton production in \pHe performed by LHCb in 2018\,\cite{LHCb-PAPER-2018-031}, the antihyperon decay contributions are addressed in a new analysis\,\cite{LHCb-PAPER-2022-006} with two complementary approaches. The dominant term, \decay{\Lbar_{\text{prompt}}}{\pbar\pip}, is reconstructed in the detector with no use of particle identification (PID) information. The ratio of the measured cross-section with the previous LHCb result gives $R_{\Lbar} = \sigma{(\decay{\pHe}{(\decay{\Lbar_{\text{prompt}}}{\pbar\pip})X)}}/\sigma(\decay{\pHe}{\bar{p}_{\text{prompt}}X})$. An inclusive approach is also followed exploiting the LHCb excellent PID performance and impact parameter resolution. All antiprotons are reconstructed, selected and distinguished between promptly and decay-produced with a template fit to the \pHe data impact parameter distribution. The ratio $R_{\Hbar} = \sigma{(\decay{\pHe}{\decay{\Hbar X}{\pbar X})}}/\sigma(\decay{\pHe}{\bar{p}_{\text{prompt}}X})$, with $\Hbar = \{\Lbar, \Sbar, \Xibarp, \Xibarz, \Omegabar\}$, is obtained by correcting the two antiproton yields by the respective efficiencies. Figure~\ref{fig:pHe_results} illustrates the $R_\Hbar$ result as a function of the transverse momentum, in excess with respect to the predictions from all theoretical models. The $R_\Lbar/R_\Hbar$ ratio is also compared to the EPOS-LHC\,\cite{EPOS-LHC} prediction, expected to be much more reliable for this quantity since it only depends on the \textit{s}-quark hadronization. The observed consistency confirms the validity of the two approaches.

\section{Conclusions}
In parallel to the study of \pp collisions, LHCb is developing a full and unique heavy-ion programme, constraining theoretical models in poorly explored kinematic regions. Since 2015, LHCb is also pioneering fixed-target physics at LHC, addressing several fields of interest beyond those of its conception. In this document, four measurements exemplifying the LHCb unique contributions in different collision systems are presented. With the ongoing and the future upgrades of the experiment, the programmes will be boosted, further widening the LHCb physics reach. LHCb is evolving more and more to a general purpose forward experiment.

\end{document}